\documentclass[preprint,showpacs,aps]{revtex4}

\usepackage{amsmath}
\usepackage{dcolumn}
\usepackage{verbatim}
\usepackage{mathrsfs}
\usepackage{graphicx}
\usepackage[usenames]{color}

\begin{document}

\title{Shell model test of quadrupole properties predicted by the rotational 
formula---the degenerate SDI interaction and non-degenerate FPD6}

\author{A. Escuderos and L. Zamick}
\affiliation{Department of Physics and Astronomy, Rutgers University,
Piscataway, New Jersey 08854}

\date{\today}

\begin{abstract} 
In the rotational model for a $K=0$ band in an even--even nucleus, there is a
single parameter---$Q_0$, the intrinsic quadrupole moment. All $B(E2)$'s in the
band and all static quadrupole moments are expressed in terms of this one 
parameter. In shell-model calculations, this does not have to be the case. In
this work, we consider ground-state bands in $^{44}$Ti, $^{46}$Ti, $^{48}$Ti,
$^{48}$Cr, and $^{50}$Cr with two different models. First, we use a Surface 
Delta
Interaction with degenerate single-particle energies (SDI-deg). We compare this
with results of a shell-model calculation using the standard interaction FPD6
and include the single-particle energy splitting. Neither model yields a 
perfect rotational $I(I+1)$ spectrum, although the SDI-deg model comes 
somewhat closer. Overall, the simple rotational formula for $B(E2)$'s and 
static quadrupole moments hangs together very nicely.
\end{abstract}


\maketitle

\section{Introduction}

In a previous publication~\cite{rezncrf06}, we considered the relationship 
between the static quadrupole moments of the $2^+_1$ states and the 
corresponding $B(E2)$'s in the $sd$ and $fp$ shells. We 
used the shell model to operationally define a ratio of intrinsic quadrupole 
moments $Q_0(S)/Q_0(B)$, where $Q_0(S)$ is obtained from the static quadrupole
moments and $Q_0(B)$, from the $B(E2)$'s. In the simple rotational model, this
ratio would equal unity, and in the harmonic vibrator model, it would equal 
zero.

There have been tests of this ratio in other models such as the Skyrme 
Hartree--Fock model by Bender et al.~\cite{bfh03,bbdh04}. This same model has
been used to test the systematics of quadrupole deformations by Jaqaman and 
Zamick~\cite{jz84}, Zheng et al.~\cite{zbz88}, Retamosa et 
al.~\cite{rupm90,prm89}, and more recently by Sagawa et al.~\cite{szz05}.
At the same time as Ref.~\cite{rezncrf06}, there appeared an article on the 
same topic but with a different approach by S.M.~Lenzi et al.~\cite{lmb06}
and, more recently, by G.~Thiamova et al.~\cite{trw06}. Very recent 
references~\cite{sbbh07,bghdgp07} show ever increasing interest in this 
subject.

In another vein, random interaction studies were performed by Vel\'azquez et
al.~\cite{vhfz03} and by Zelevinsky and Volya~\cite{zv04}. They found two 
spikes (i.e., high probabilities) in the Alaga ratio $A=5 Q^2/[16\pi B(E2)_{0 
\rightarrow 2}]$ at $A=0$ and $A=A_0=4/49$, which can be 
associated with the vibrational and rotational limits, respectively.

On the other hand, for the nuclei that we considered in Ref.~\cite{rezncrf06},
the experimental ratio $Q_0(S)/Q_0(B)$
was, for the most part, large, sometimes exceeding one, e.g., for $^{20}$Ne and
$^{50}$Cr. The one exception was $^{40}$Ar, where the ratio was $0.06$.

We would like to mention that Poves et al.~\cite{czpm94} have shown that, in a
full $fp$ space, one gets a quasi-rotational band in $^{48}$Cr. The spectrum 
is not $I(I+1)$ exactly but there are strong intraband $E2$ transitions 
between the levels.

In this work, we shall extend this study by considering states of higher 
angular momentum as well. All calculations have been done using the 
shell-model code ANTOINE~\cite{cn99}.

We define
\begin{equation}
R_{SB}= \frac{Q_0(S)_{I=2}}{Q_0(B)_{2\rightarrow 0}}=
-\frac{7}{2\sqrt{16 \pi}} \frac{Q(2^+)}{\sqrt{B(E2)_{2\rightarrow 0}}} =
-0.4936659 \frac{Q(2^+)}{\sqrt{B(E2)_{2\rightarrow 0}}} \ ,
\end{equation}
and
\begin{equation}
M(Q)_I=\frac{Q_0(S)_I}{Q_0(S)_2}=\frac{2}{7} \frac{2I+3}{I} \frac{Q(I)}{Q(2)}
\ ,
\end{equation}
where $Q(I)$ is the (laboratory) static quadrupole moment of a state of angular
momentum $I$. Note that the Alaga ratio mentioned above is 
$A=\frac{4}{49}|R_{SB}(2)|^2$.

We also define
\begin{equation}
M(B)_{I\rightarrow I-2}=\frac{Q_0(B)_{I\rightarrow I-2}}{Q_0(B)_{2\rightarrow 
0}} \ .
\end{equation}

Now for a $K=0$ rotational band, we have
\begin{equation}
B(E2)_{I\rightarrow I-2}=\frac{5}{16 \pi} (I 2 0 0 | I-2,0)^2 Q^2_B(I) \ ,
\end{equation}
\begin{equation}
B(E2)_{2\rightarrow 0}=\frac{5}{16\pi} \frac{Q^2_B(2)}{5} \ .
\end{equation}
Hence, 
\begin{equation}
M(B)^2_{I \rightarrow I-2}=\frac{2}{15} \frac{(2I-1)(2I+1)}{I(I-1)} 
\frac{B(E2)_{I\rightarrow I-2}}{B(E2)_{2\rightarrow 0}} \ .
\end{equation}
Note that both $M(Q)_2$ and $M(B)_{2\rightarrow 0}$ are equal to 1 by 
definition.

\section{Preliminary remarks}

\subsection{The Schematic Surface Delta Interaction}

In the past, schematic models, despite giving somewhat oversimplified 
descriptions of the structures of nuclei, proved to be invaluable in casting 
insights into the trends of nuclear structure. As an example, Elliott's SU(3) 
model showed how one could approach rotational-model spectra in the Shell 
Model~\cite{e58}. This was shown with a two-body momentum-dependant long-range 
quadrupole-quadrupole interaction and did not include the effects of spin-orbit
splitting. Indeed, SU(3) models emanating from this interaction are emphasized
in Refs.~\cite{lmb06,trw06}.

In this work, the main thrust will be to  use a realistic interaction with 
correct single particle splittings. However, since our results for $R_{SB}$
agree with the rotational model even when the spectra are not rotational, we 
are motivated to get insight into this result by using a different schematic
interaction, one which does not have $I(I+1)$ spectra but still exhibit 
collective behaviour. In contrast to Elliott's long range 
interaction~\cite{e58}, we will use the surface delta interaction of 
Moszkowski with degenerate single particle energies~\cite{gm65}.

\begin{figure}[ht]
\includegraphics[scale=.5]{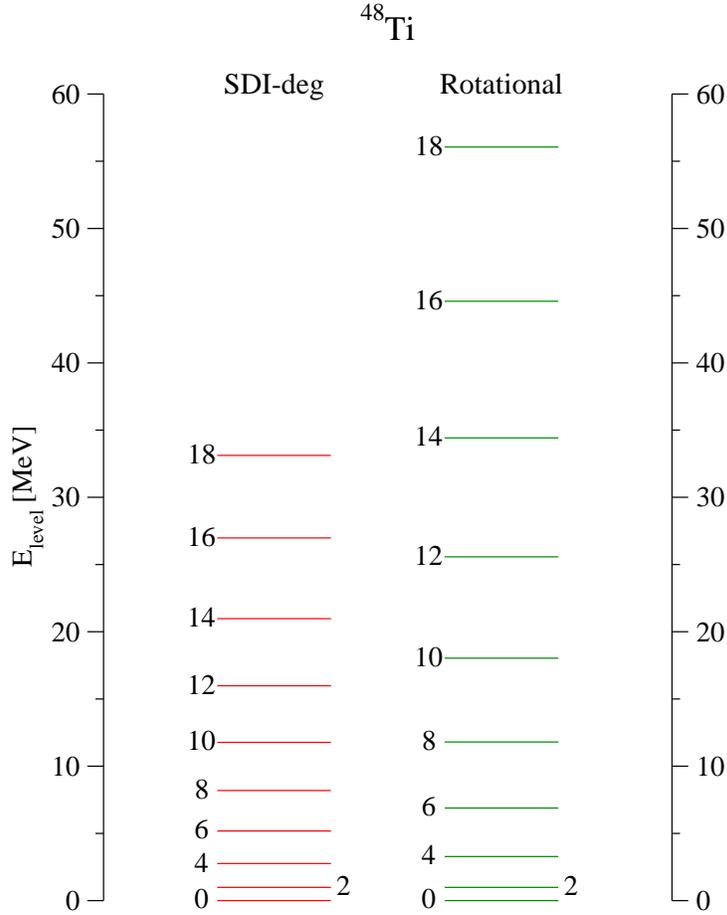}
\caption{Comparison of the SDI-deg and rotational $I(I+1)$ spectra for 
$^{48}$Ti. Both interactions have been fitted to reproduce the experimental
splitting of the $0^+_1$ and $2^+_1$ states.} \label{fig:rot-sdi}
\end{figure}

In Fig.~\ref{fig:rot-sdi}, we show the calculated spectrum of $^{48}$Ti  with a
surface delta interaction as compared with the $I(I+1)$ rotational spectrum. 
The parameters have been adjusted so that the excitation energies of the first 
$2^+$ states are the same. The rotational spectrum is more spread out but the 
surface delta spectrum still has rotational features and the spectrum is 
actually closer to the truth than the $I(I+1)$, at least for $^{48}$Ti. Hence, 
the surface delta interaction will serve as a good counterpoint to the more 
realistic interaction considered in the next section.

\subsection{Experimental Comparison}

In Ref.~\cite{lmb06}, S.M.~Lenzi et al. compare the $B(E2)$'s of $N=Z$ nuclei, 
$^{44}$Ti and $^{48}$Cr, with the SU(3) model of Elliott~\cite{e58}, making use of the
momentum-dependent quadrupole-quadrupole interaction. Although the SU(3) model gives a
rotational spectrum, the $BE(2)$'s do not follow the rotational formula.
Rather, with increasing $J$, they decrease relative to the rotational formula.
The authors make the point that, despite the absence of spin-orbit splitting,
the SU(3) results are not too bad compared with experiment. For $^{48}$Cr they refer to 
the experiments of Brandolini et al.~\cite{betal98}. We shall also use this reference for our 
analysis of $Q_0(B)_{I\rightarrow I-2}/Q_0(B)_{2\rightarrow 0}$.

In tables~\ref{tab:48cr-ex} and \ref{tab:50cr-ex} we can see the results from 
Ref.~\cite{betal98} for $^{48}$Cr and $^{50}$Cr, respectively. Note that, in the case of
$^{48}$Cr, $M(B)_{I\rightarrow I-2}$ decreases almost linearly with $I$. There is
more experimental information, but, beyond the results shown in the tables, 
things get lost in band crossing.

\begin{table}[ht]
\caption{Experimental $B(E2)$'s [e$^2$~fm$^4$] and ratio $M(B)_{I \to I-2}$ 
(see text) for $^{48}$Cr; experimental data are taken from 
Ref.~\cite{betal98}, except the $B(E2)_{2 \to 0}$, which is taken from 
Ref.~\cite{rnt01}.} \label{tab:48cr-ex}
\begin{tabular*}{.65\textwidth}[t]{@{\extracolsep{\fill}}ccc}
\toprule
 & $B(E2)_\text{exp}$ [e$^2$ fm$^4$] & $M(B)_{I\rightarrow I-2}$ \\ \colrule
$2 \to 0$ & 272\hphantom{(000)} & 1.00000 \\
$4 \to 2$ & 329(110) & 0.92016 \\
$6 \to 4$ & 301(78)\hphantom{0} & 0.83863 \\
$8 \to 6$ & 230(69)\hphantom{0} & 0.71651 \\
$10 \to 8$ & 195(54)\hphantom{0} & 0.65098 \\
$12 \to 10$ & 167(25)\hphantom{0} & 0.59716 \\
$14 \to 12$ & 105(18)\hphantom{0} & 0.47057 \\
$16 \to 14$ & \hphantom{0}37(8)\hphantom{00} & 0.27805 \\
\botrule
\end{tabular*}
\end{table}

\begin{table}[ht]
\caption{Same as Table~\ref{tab:48cr-ex} but for $^{50}$Cr.} 
\label{tab:50cr-ex}
\begin{tabular*}{.65\textwidth}[t]{@{\extracolsep{\fill}}ccc}
\toprule
 & $B(E2)_\text{exp}$ [e$^2$ fm$^4$] & $M(B)_{I\rightarrow I-2}$ \\ \colrule
$2 \to 0$ & 216\hphantom{(00)} & 1.00000 \\
$4 \to 2$ & 204(57) & 0.81309 \\
$6 \to 4$ & 235(47) & 0.83154 \\
$8 \to 6$ & 205(51) & 0.75909 \\
$10(1) \to 8$ & \hphantom{0}72(14) & 0.44389 \\
$10(2) \to 8$ & 131(26) & 0.59875 \\
\botrule
\end{tabular*}
\end{table}

\section{Results}

As in Ref.~\cite{rezncrf06}, in Table~\ref{tab:rsb} we check
the relationship between the $B(E2)_{2^+_1\rightarrow 0^+_1}$ and the static
quadrupole moment $Q(2^+_1)$.

\begin{table}[ht]
\caption{$Q(2^+)$ [$e$ fm$^2$], $B(E2)_{2^+ \rightarrow 0^+}$ [$e^2$ fm$^4$], 
and $R_{SB}$ for $^{44,46,48}$Ti and $^{48,50}$Cr obtained from a full $fp$ 
shell calculation with the interactions SDI and FPD6.} \label{tab:rsb}
\begin{tabular*}{.9\textwidth}[t]{@{\extracolsep{\fill}}cddrddr}
\toprule
 & \multicolumn{3}{c}{SDI-deg} & \multicolumn{3}{c}{FPD6} \\ 
\cline{2-4} \cline{5-7}
 & \multicolumn{1}{c}{$Q(2^+)$} &
 \multicolumn{1}{c}{$B(E2)\downarrow$} & 
 \multicolumn{1}{c}{$R_{SB}$} & 
 \multicolumn{1}{c}{$Q(2^+)$} & 
 \multicolumn{1}{c}{$B(E2)\downarrow$} & 
 \multicolumn{1}{c}{$R_{SB}$} \\
\colrule
$^{44}$Ti & -26.319 & 165.74 & 1.0092 & -20.156 & 121.45 & 0.9029 \\
$^{46}$Ti & -29.349 & 206.06 & 1.0093 & -22.071 & 136.41 & 0.9329 \\
$^{48}$Ti & -32.337 & 247.38 & 1.0149 & -17.714 & 112.16 & 0.8257 \\
$^{48}$Cr & -37.936 & 377.60 & 0.9638 & -33.271 & 275.68 & 0.9892 \\
$^{50}$Cr & -41.681 & 435.82 & 0.9856 & -30.955 & 243.80 & 0.9787 \\
\botrule
\end{tabular*}
\end{table}

In the rotational model,
\begin{equation}
Q(2^+_1)=-\frac{2}{7} Q_0(S) \ ,
\end{equation}
\begin{equation}
B(E2)_{2^+_1\rightarrow 0^+_1}=\frac{Q_0(B)^2}{16\pi} \ .
\end{equation}
Furthermore, the ratio $R_{SB}=Q_0(S)/Q_0(B)$ should equal 1.

\begin{figure}[ht]
\includegraphics[scale=.5]{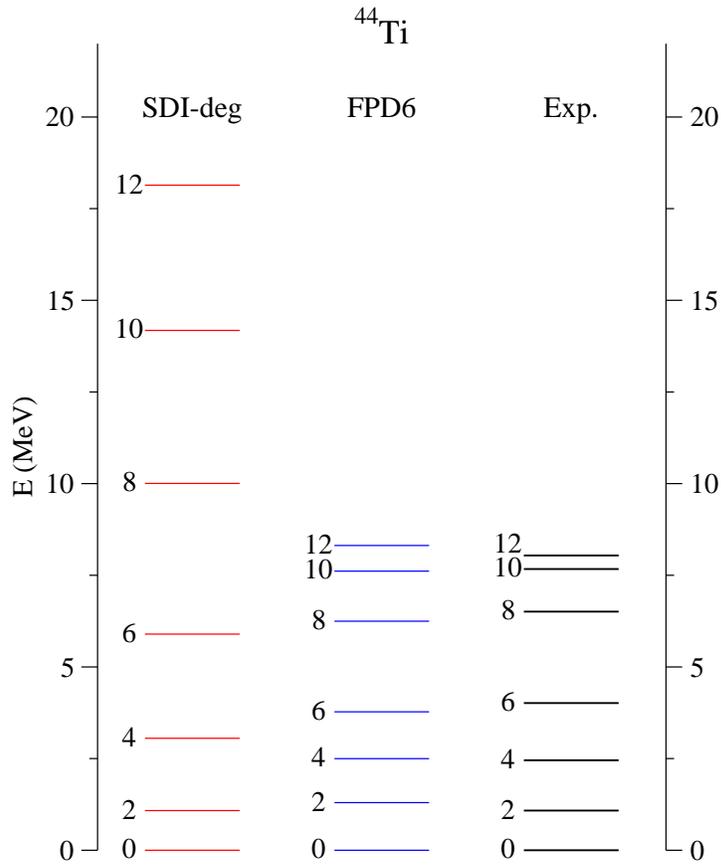}
\caption{Full $fp$ space calculations of even-$J$ states in $^{44}$Ti with the
SDI-deg and FPD6 interactions, and comparison with experiment.} 
\label{fig:ti44}
\end{figure}
\begin{figure}[ht]
\includegraphics[scale=.5]{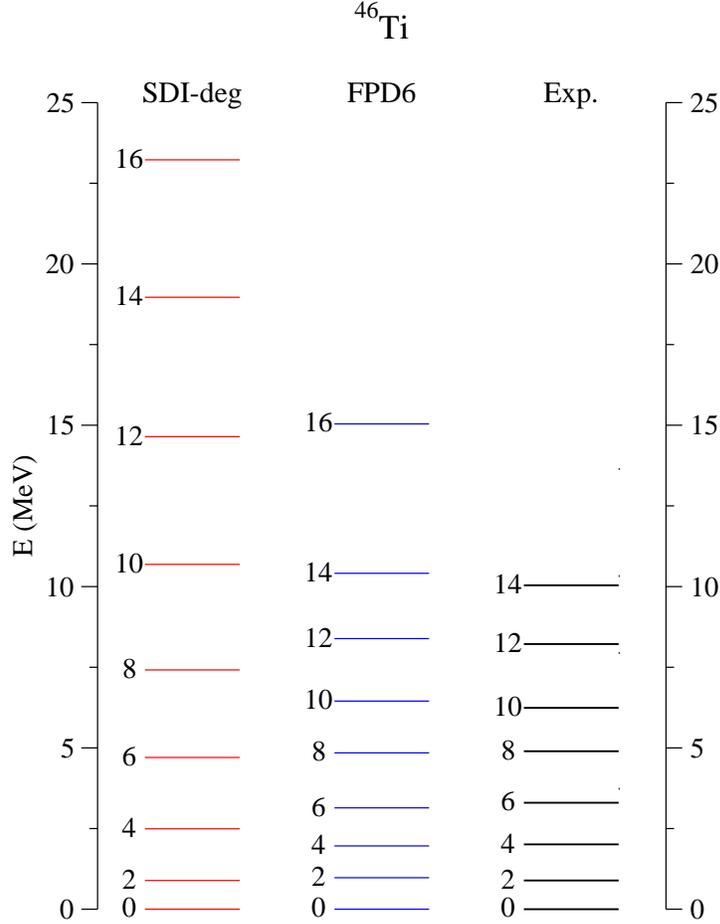}
\caption{The same as Fig.~\ref{fig:ti44} for $^{46}$Ti.} \label{fig:ti46}
\end{figure}
\begin{figure}[ht]
\includegraphics[scale=.5]{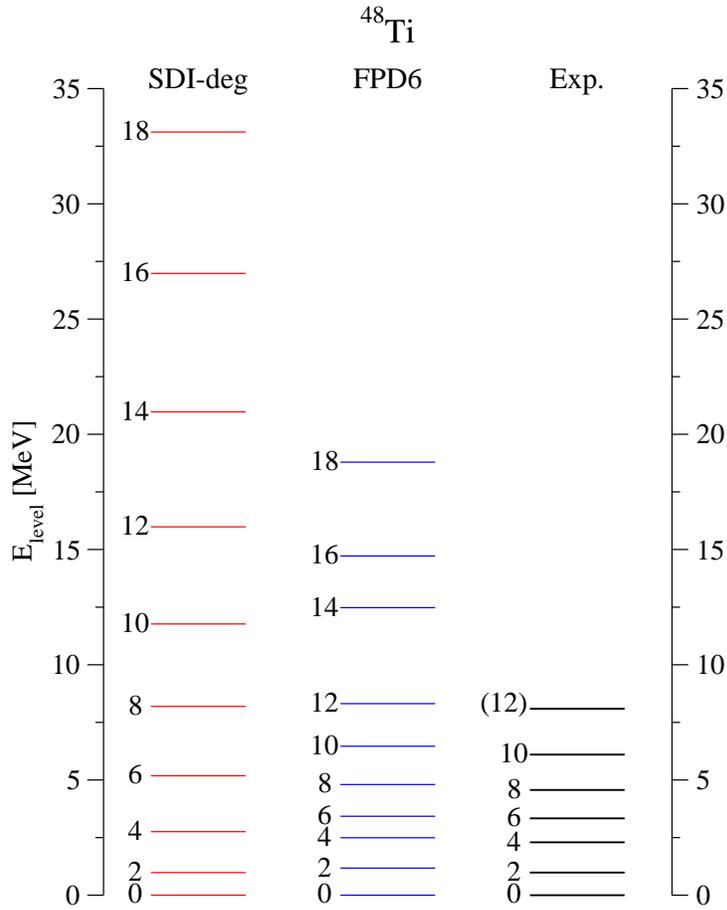}
\caption{The same as Fig.~\ref{fig:ti44} for $^{48}$Ti.} \label{fig:ti48}
\end{figure}
\begin{figure}[ht]
\includegraphics[scale=.5]{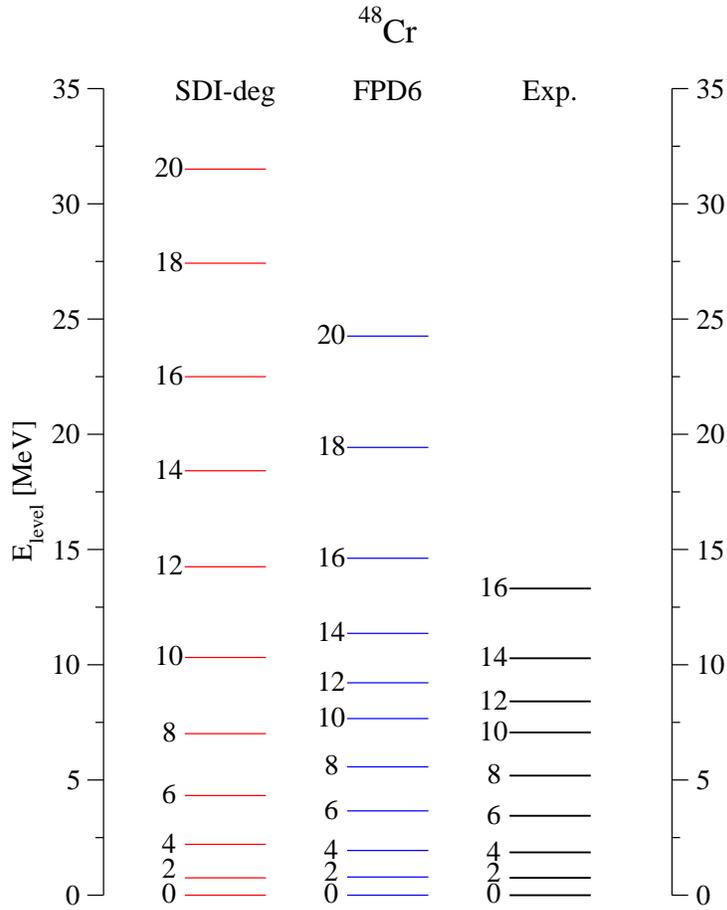}
\caption{The same as Fig.~\ref{fig:ti44} for $^{48}$Cr.} \label{fig:cr48}
\end{figure}
\begin{figure}[ht]
\includegraphics[scale=.5]{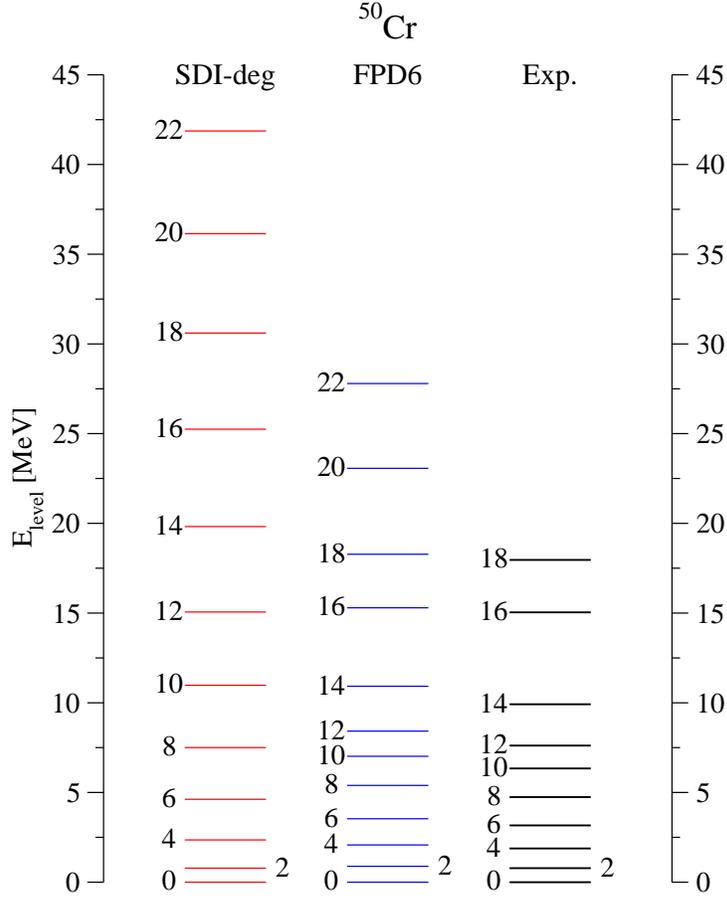}
\caption{The same as Fig.~\ref{fig:ti44} for $^{50}$Cr.} \label{fig:cr50}
\end{figure}

In Table~\ref{tab:rsb} we give $R_{SB}$ for two models: a surface delta 
interaction with degenerate single-particle energies (SDI-deg) and the more 
realistic calculation with the FPD6 interaction including single-particle 
energy splittings, both in a full $fp$ space. The strength of the SDI-deg 
interaction was chosen to fit the experimental excitation energy of the $2^+_1$
state. The nuclei considered 
are $^{44}$Ti, $^{46}$Ti, $^{48}$Ti, $^{48}$Cr, and $^{50}$Cr. Note that, from
Figs.~\ref{fig:ti44}--\ref{fig:cr50}, neither SDI-deg or FPD6 have 
pure rotational spectra, although SDI is closer to a rotational spectrum, 
undoubtedly due to the fact that there are no single-particle splittings.

\begin{table}[ht]
\caption{Ratios of intrinsic quadrupole moments $M(Q)_I$ and 
$M(B)_{I\rightarrow I-2}$ (see text) for $^{44}$Ti using both SDI-deg and FPD6 
interactions.} 
\label{tab:44ti}
\begin{tabular*}{.8\textwidth}[t]{@{\extracolsep{\fill}}cdddd}
\toprule
 & \multicolumn{2}{c}{SDI-deg} & \multicolumn{2}{c}{FPD6} \\ 
\cline{2-3} \cline{4-5}
$I\rightarrow I-2$ & \multicolumn{1}{c}{$M(Q)_I$} & 
\multicolumn{1}{c}{$M(B)_{I\rightarrow I-2}$} & \multicolumn{1}{c}{$M(Q)_I$} & 
\multicolumn{1}{c}{$M(B)_{I\rightarrow I-2}$} \\
\colrule
$2\rightarrow 0$ & 1 & 1 & 1 & 1 \\
$4\rightarrow 2$ & 1.0092 & 0.9741 & 1.0534 & 0.9764 \\
$6\rightarrow 4$ & 1.0176 & 0.9096 & 1.1047 & 0.8554 \\
$8\rightarrow 6$ & 1.0276 & 0.8241 & 0.8498 & 0.6971 \\
$10\rightarrow 8$ & 1.0406 & 0.7239 & 0.7800 & 0.6802 \\
$12\rightarrow 10$ & 1.0556 & 0.5498 & 0.8472 & 0.5130 \\
\botrule
\end{tabular*}
\end{table}

The SDI-deg results for $R_{SB}$ are remarkably close to unity, ranging from 
0.964 to 1.009. With FPD6, the results range from 0.826 to 0.989---somewhat 
farther from unity, but again noticeably close. Remember that, in a very simple
model for a vibrational nucleus, the ratio of $R_{SB}$ would be zero. Clearly 
the shell model shows greater resistence for the intrinsic quadrupole moment 
$Q_0(S)$ to become small, as we might be led to believe from collective 
arguments.

\begin{table}[ht]
\caption{The same as Table~\ref{tab:44ti} for $^{46}$Ti.} \label{tab:46ti}
\begin{tabular*}{.8\textwidth}[t]{@{\extracolsep{\fill}}cdddd}
\toprule
 & \multicolumn{2}{c}{SDI-deg} & \multicolumn{2}{c}{FPD6} \\ 
\cline{2-3} \cline{4-5}
$I\rightarrow I-2$ & \multicolumn{1}{c}{$M(Q)_I$} & 
\multicolumn{1}{c}{$M(B)_{I\rightarrow I-2}$} & \multicolumn{1}{c}{$M(Q)_I$} & 
\multicolumn{1}{c}{$M(B)_{I\rightarrow I-2}$} \\
\colrule
$2\rightarrow 0$ & 1 & 1 & 1 & 1 \\
$4\rightarrow 2$ & 1.0159 & 0.9878 & 1.0349 & 0.9975 \\
$6\rightarrow 4$ & 1.0328 & 0.9351 & 0.9483 & 0.9398 \\
$8\rightarrow 6$ & 1.0295 & 0.8152 & 0.9522 & 0.8626 \\
$10\rightarrow 8$ & 1.0270 & 0.7562 & 0.9362 & 0.7436 \\
$12_1\rightarrow 10$ & 1.0131 & 0.0044 & 0.6763 & 0.4493 \\
$12_2\rightarrow 10$ & 1.0658 & 0.6654 & 0.2357 & 0.2245 \\
$14\rightarrow 12_1$ & 1.0331 & 0.6803 & 0.6724 & 0.3804 \\
$14\rightarrow 12_2$ & \multicolumn{1}{c}{$\sf ~~~~~ ''$} & 0.0007 & 
 \multicolumn{1}{c}{$\sf ~~~~~ ''$} & 0.2709 \\
$16\rightarrow 14$ & 1.0480 & 0.5212 & 0.8911 & 0.0644 \\
\botrule
\end{tabular*}
\end{table}

In Tables~\ref{tab:44ti}--\ref{tab:50cr}, we show the results of $M(Q)_I$ and
$M(B)_{I\rightarrow I-2}$ for all the nuclei with both interactions.
We first look at $M(Q)_I$ for the surface delta interaction. For $^{44}$Ti, 
$^{46}$Ti, and $^{48}$Ti, the values are slightly larger but remarkably close
to unity, even for very high spins, e.g., $I=12, 16$, and $18$ for $A=44,46$,
and $48$, respectively. In $^{48}$Cr the results are not so close beyond 
$I=6$, the values 
being 0.859, 0.766, and 0.779 for $I=8,10$, and $12_1$. Strangely, for 
$^{50}$Cr the results are better up to $I=10$. All in all, though, we are very
far away from the simple vibrational limit of zero.

\begin{table}[ht]
\caption{The same as Table~\ref{tab:44ti} for $^{48}$Ti.} \label{tab:48ti}
\begin{tabular*}{.8\textwidth}[t]{@{\extracolsep{\fill}}cdddd}
\toprule
 & \multicolumn{2}{c}{SDI-deg} & \multicolumn{2}{c}{FPD6} \\ 
\cline{2-3} \cline{4-5}
$I\rightarrow I-2$ & \multicolumn{1}{c}{$M(Q)_I$} & 
\multicolumn{1}{c}{$M(B)_{I\rightarrow I-2}$} & \multicolumn{1}{c}{$M(Q)_I$} & 
\multicolumn{1}{c}{$M(B)_{I\rightarrow I-2}$} \\
\colrule
$2\rightarrow 0$ & 1 & 1 & 1 & 1 \\
$4\rightarrow 2$ & 1.0172 & 1.0015 & 0.8654 & 1.0310 \\
$6_1\rightarrow 4$ & 1.0276 & 0.9884 & -0.8738 & 0.5016 \\
$6_2\rightarrow 4$ & 0.2782 & 0.0309 & 0.8041 & 0.7124 \\
$8\rightarrow 6_1$ & 1.0319 & 0.9598 & 0.6134 & 0.5768 \\
$8\rightarrow 6_2$ & \multicolumn{1}{c}{$\sf ~~~~~ ''$} & 0.0016 & 
 \multicolumn{1}{c}{$\sf ~~~~~ ''$} & 0.3832 \\
$10\rightarrow 8$ & 1.0340 & 0.9193 & 0.8153 & 0.5682 \\
$12\rightarrow 10$ & 1.0347 & 0.8589 & 0.5462 & 0.3616 \\
$14\rightarrow 12$ & 1.0392 & 0.7821 & 0.8447 & 0.1491 \\
$16\rightarrow 14$ & 1.0519 & 0.6657 & 0.9128 & 0.4386 \\
$18\rightarrow 16$ & 1.0614 & 0.5062 & 1.0951 & 0.0646 \\
\botrule
\end{tabular*}
\end{table}

If we look at $M(B)_{I\rightarrow I-2}$ with SDI-deg, the results up to the 
transition $8 \rightarrow 6$ are all greater than 0.8 and less than unity for 
all nuclei considered. Beyond that, there are some lower results that may be 
due to band crossing, e.g., the value of $M(B)_{12_1 \rightarrow 10_1}$ in 
$^{46}$Ti is 0.0044; however, $M(B)_{12_2 \rightarrow 10_1}$ is 0.6654.

\begin{table}[ht]
\caption{The same as Table~\ref{tab:44ti} for $^{48}$Cr.} \label{tab:48cr}
\begin{tabular*}{.8\textwidth}[t]{@{\extracolsep{\fill}}cdddd}
\toprule
 & \multicolumn{2}{c}{SDI-deg} & \multicolumn{2}{c}{FPD6} \\ 
\cline{2-3} \cline{4-5}
$I\rightarrow I-2$ & \multicolumn{1}{c}{$M(Q)_I$} & 
\multicolumn{1}{c}{$M(B)_{I\rightarrow I-2}$} & \multicolumn{1}{c}{$M(Q)_I$} & 
\multicolumn{1}{c}{$M(B)_{I\rightarrow I-2}$} \\
\colrule
$2\rightarrow 0$ & 1 & 1 & 1 & 1 \\
$4\rightarrow 2$ & 0.9797 & 0.9998 & 1.0088 & 0.9887 \\
$6\rightarrow 4$ & 0.9028 & 0.9851 & 0.9679 & 0.9600 \\
$8\rightarrow 6$ & 0.8586 & 0.9592 & 0.9370 & 0.9101 \\
$10\rightarrow 8$ & 0.7660 & 0.9121 & 0.7703 & 0.8028 \\
$12_1\rightarrow 10$ & 0.7788 & 0.8839 & 0.1381 & 0.5277 \\
$12_2\rightarrow 10$ & 1.3175 & 0.0985 & 0.5786 & 0.5102 \\
$14_1\rightarrow 12_1$ & 1.3722 & 0.0790 & 0.1675 & 0.5024 \\
$14_2\rightarrow 12_1$ & 0.7588 & 0.8642 & 0.6645 & 0.1394 \\
$16_1\rightarrow 14_1$ & 1.4145 & 0.4311 & 0.1537 & 0.3536 \\
$16_2\rightarrow 14_1$ & 0.7379 & 0.1255 & 0.6231 & 0.1401 \\
$16_2\rightarrow 14_2$ & \multicolumn{1}{c}{$\sf ~~~~~ ''$} & 0.6032 & 
 \multicolumn{1}{c}{$\sf ~~~~~ ''$} & 0.5163 \\
$18_1\rightarrow 16_1$ & 1.1348 & 0.0000 & 0.5938 & 0.0577 \\
$18_1\rightarrow 16_2$ & \multicolumn{1}{c}{$\sf ~~~~~ ''$} & 0.0000 & 
 \multicolumn{1}{c}{$\sf ~~~~~ ''$} & 0.4369 \\
$18_2\rightarrow 16_1$ & 0.8025 & 0.0000 & 0.7394 & 0.0157 \\
$18_2\rightarrow 16_2$ & \multicolumn{1}{c}{$\sf ~~~~~ ''$} & 0.5765 &
 \multicolumn{1}{c}{$\sf ~~~~~ ''$} & 0.1430 \\
$20\rightarrow 18$ & 1.0753 & 0.4473 & 0.8097 & 0.1183 \\
\botrule
\end{tabular*}
\end{table}

Results for the more realistic FPD6 interaction including single-particle
energies are also shown in Tables~\ref{tab:44ti}--\ref{tab:50cr}. The results
are over all not as close to unity as with SDI-deg. Still, one gets some 
substantial static quadrupole moments. Sometimes, the lowest state of a given
angular momentum does not belong to a $K=0$ band, e.g., the $I=6^+_1$ state in
$^{48}$Ti or the $10^+_1$ state in $^{50}$Cr. The sign of the static quadrupole
moment is opposite to what one would get assuming $K=0$. In $^{48}$Ti, the 
second $6^+$ state fits better into this profile. The near degeneracies of the 
$6^+_1$ and $6^+_2$ states in $^{48}$Ti has been discussed 
previously~\cite{rz01}.

\begin{table}[ht]
\caption{The same as Table~\ref{tab:44ti} for $^{50}$Cr.} \label{tab:50cr}
\begin{tabular*}{.8\textwidth}[t]{@{\extracolsep{\fill}}cdddd}
\toprule
 & \multicolumn{2}{c}{SDI-deg} & \multicolumn{2}{c}{FPD6} \\ 
\cline{2-3} \cline{4-5}
$I\rightarrow I-2$ & \multicolumn{1}{c}{$M(Q)_I$} & 
\multicolumn{1}{c}{$M(B)_{I\rightarrow I-2}$} & \multicolumn{1}{c}{$M(Q)_I$} & 
\multicolumn{1}{c}{$M(B)_{I\rightarrow I-2}$} \\
\colrule
$2\rightarrow 0$ & 1 & 1 & 1 & 1 \\
$4\rightarrow 2$ & 0.9881 & 1.0027 & 1.0014 & 1.0074 \\
$6\rightarrow 4$ & 0.9631 & 0.9954 & 0.5587 & 0.8785 \\
$8\rightarrow 6$ & 0.9332 & 0.9807 & 0.6464 & 0.8154 \\
$10_1\rightarrow 8$ & 0.9064 & 0.9557 & -0.9601 & 0.2678 \\
$10_2\rightarrow 8$ & 1.0923 & 0.0094 & 0.2300 & 0.6599 \\
$12_1\rightarrow 10_1$ & 0.8857 & 0.9751 & -0.3819 & 0.3115 \\
$12_1\rightarrow 10_2$ & \multicolumn{1}{c}{$\sf ~~~~~ ''$} & 0.0959 &
 \multicolumn{1}{c}{$\sf ~~~~~ ''$} & 0.2502 \\
$12_2\rightarrow 10_1$ & 1.1533 & 0.0058 & 0.5044 & 0.1502 \\
$12_2\rightarrow 10_2$ & \multicolumn{1}{c}{$\sf ~~~~~ ''$} & 0.8039 &
 \multicolumn{1}{c}{$\sf ~~~~~ ''$} & 0.5064 \\
$14_1\rightarrow 12_1$ & 0.8482 & 0.8672 & -0.2057 & 0.3662 \\
$14_1\rightarrow 12_2$ & \multicolumn{1}{c}{$\sf ~~~~~ ''$} & 0.1310 &
 \multicolumn{1}{c}{$\sf ~~~~~ ''$} & 0.1013 \\
$14_2\rightarrow 12_1$ & 1.1224 & 0.0131 & 0.3175 & 0.0947 \\
$14_2\rightarrow 12_2$ & \multicolumn{1}{c}{$\sf ~~~~~ ''$} & 0.1404 &
 \multicolumn{1}{c}{$\sf ~~~~~ ''$} & 0.4982 \\
$16\rightarrow 14_1$ & 0.8832 & 0.8584 & 0.2218 & 0.0943 \\
$16\rightarrow 14_2$ & \multicolumn{1}{c}{$\sf ~~~~~ ''$} & 0.0161 &
 \multicolumn{1}{c}{$\sf ~~~~~ ''$} & 0.5892 \\
$18\rightarrow 16$ & 1.1256 & 0.0266 & 0.1888 & 0.4213 \\
$20\rightarrow 18$ & 1.1306 & 0.6272 & 0.5944 & 0.0619 \\
$22\rightarrow 20$ & 1.0694 & 0.4435 & 0.9263 & 0.1189 \\
\botrule
\end{tabular*}
\end{table}

In previous works on $^{50}$Cr~\cite{zfz96,zz96}, it was noted that the first
$10^+$ state did not belong to the $K=0$ ground state band. Indeed, if one must
choose a $K$ value, it would seem $K=10$ is the best for the $10_1^+$ state. 
This is supported by the fact that the static quadrupole moment is large and 
positive, while the $K=0$ static moments are negative. Moreover, if it were 
strictly $K=10$ and the $8^+_1$ were strictly $K=0$, the $B(E2)$ would be 
strongly inhibited. The small value of $M(B)_{10_1\rightarrow 8}=0.2678$ 
somewhat supports this. The decay $10^+_2 \rightarrow 8^+_1$ is stronger, with 
$M(B)_{10_2\rightarrow 8}=0.6599$.

Let us briefly discuss Figs.~\ref{fig:ti44}--\ref{fig:cr50} corresponding to
$^{44}$Ti, $^{46}$Ti, $^{48}$Ti, $^{48}$Cr, and $^{50}$Cr, respectively.
Although not perfect, the FPD6 interaction in a full $fp$ space yields a pretty
good agreement for the energy levels of all nuclei here considered. The SDI 
interaction, for which the 0--2 splitting is fitted to experiment, gives a
more spread out spectrum. It is closer to an $I(I+1)$ spectrum than results
with FPD6, but still significantly different. The spreading of the spectrum 
with SDI-deg is mainly due to the fact that there are no single-particle 
splittings in this model, i.e., $\epsilon_{f_{7/2}}=\epsilon_{f_{5/2}}=
\epsilon_{p_{3/2}}=\epsilon_{p_{1/2}}$.

Just to give some numbers, in $^{48}$Ti the experimental energy of the $12^+_1$
state is 8.09~MeV, FPD6 gives 8.31~MeV, and SDI-deg gives 15.98~MeV. Using the
simple rotational model and fitting the 0--2 splitting to experiment, the 
$12^+_1$ state would, with an $I(I+1)$ spectrum, be at 25.57~MeV.

\section{The harmonic vibrator}

In the harmonic vibrator model, the nuclear shape oscillates between oblate and
prolate. One gets equally spaced spectra, i.e., the ground state has angular
momentum $I=0$, the first excited state has angular momentum $I=2$ and energy
$E(2)$. At $2 E(2)$ there are states with $I=0,2$, and 4; at $3 E(2)$ there are
states with $I=0,2,3,4$, and 6; etc.

The selection rules and $B(E2)$'s relations are given in Bohr and Mottelson
vol.~2, page 349~\cite{bm75}. The equations for a transition are
\begin{equation}
\sum_{\zeta_{n-1}I_{n-1}} B(E\lambda;n_\lambda \zeta_n I_n \rightarrow 
n_{\lambda -1} \zeta_{n-1} I_{n-1}) = n_\lambda B(E\lambda, n_\lambda =1 
\rightarrow n_\lambda =0) \ ,
\end{equation}
i.e., there is stimulated emission---the more quanta there are, the bigger the
$B(E\lambda)$. In the above, $n_\lambda$ is the number of vibrational quanta,
$I_n$ is the angular momentum and $\zeta_n$ stands for any additional quantum
numbers. Bohr and Mottelson also give the selection rule $\Delta n_\lambda = 
\pm 1$, which implies that static quadrupole moments vanish, consistent with 
the opening sentence in this section~\cite{bm75}. 

Since we are considering transitions from a state with the maximum angular
momentum ($I_{\text{max}}$) of all degenerate states with $n_\lambda$ quanta, 
the final state for $E2$ transitions must have $I=I_{\text{max}}-2$, so there 
is only one term in the left-hand side: $I_{\text{max}}-2$. Note that 
$n_\lambda = I_{\text{max}}/2$
\begin{equation}
B(E2)_{I \rightarrow I-2} = \frac{I}{2} B(E2)_{2 \rightarrow 0} \ .
\end{equation}
Hence, we obtain
\begin{equation}
M_\text{vib}(B)_{I\rightarrow I-2} = \left( \frac{1}{15} 
\frac{(2I-1)(2I+1)}{I-1} \right)^{1/2} \ .
\end{equation}
Note that $M(B)_{I\rightarrow I-2}$ increases steadily with $I$, which is not 
the case with SDI-deg or FPD6. Some values of $M_\text{vib}(B)_{I\rightarrow 
I-2}$ for $I=0,2,4,6,8$, and 10 are, respectively, $1,1.183,1.381,1.558,1.719$,
and 1.867. For large values of $I$, we have
\begin{equation}
M(B)_{I\rightarrow I-2} \rightarrow \left( \frac{4}{15} I \right)^{1/2} \ ,
\end{equation}
still growing steadily with $I$.

As mentioned before, the vibrational prediction for static quadrupole moments
is $M_\text{vib}(Q)_I=0$, which is certainly not the case with SDI-deg or FPD6.


\section{Closing remarks}

This work is an extension of previous work by Robinson et~al~\cite{rezncrf06},
where it was noted that, for a large variety of nuclei, the simple rotational
formula, if fitted to the experimental $B(E2)$ from the ground state to the
$2_1^+$ state, could give a good result for the static quadruple moment of the
$2_1^+$ state, i.e., in the notation of this work, $R_{SB}$ is close to one. 
The one exception is $^{40}$Ar.

In this work, we extend  the calculations to higher energy and  higher angular
momentum. We use both a phenomenological interaction FPD6 with realistic 
single-particle splittings, and, as a counterpoint to $Q\cdot Q$, we use the 
schematic surface delta interaction. We claim the latter shows also collective
properties. As seen in Fig.~\ref{fig:rot-sdi}, its spectrum is not rotational 
but it does have features that many nuclei possess---a collective appearance; 
and it should be noted that the nuclei we consider do not have rotational 
spectra either. We do not include single-particle splitings in SDI, and this is
the main reason, rather than the SDI per se, that the spectrum, though less 
spread out than in the rotational model, is more spread out than with a 
realistic interaction or, indeed, experiment. The effects of single-particle 
splittings are taken care of in the realistic case.

The results with SDI for $M(Q)_I$ (ratios of intrinsic quadrupole moments) is 
close to one for many nuclei and many angular momenta $I$. One only runs into 
trouble when one has near degeneracies like $6_1^+$ and $6_2^+$ in $^{48}$Ti, 
as well as $10_1^+$ and $10_2^+$ in $^{50}$Cr. These degeneracies have been 
addressed previously by Zamick et~al.~\cite{rz01,zfz96,zz96}. And when one 
gets band crossings, the situation can get confused.

But still, all in all, with SDI we get some remarkable agreements with the
rotational formulas for ratios between static quadrupole moments and
$B(E2)$'s, not only for angular momentum $I=2$, but for higher $I$ as well.
The realistic interactions also yield similar results, although perhaps not
quite as definitive as does the schematic SDI.

\begin{acknowledgments}
We thank Steve Moszkowski for his comments and interest. A.E. acknowledges
support from the Secretar\'{\i}a de Estado de Educaci\'on y Universidades 
(Spain) and the European Social Fund.
\end{acknowledgments}

\end{document}